\documentstyle[12pt]{article}
\begin{document}
\def\beq{\begin{equation}}
\def\eeq{\end{equation}}
\def\bea{\begin{eqnarray}}
\def\eea{\end{eqnarray}}
\def\ve{\vert}
\def\vel{\left|}
\def\ver{\right|}
\def\nnb{\nonumber}
\def\ga{\left(}
\def\dr{\right)}
\def\aga{\left\{}
\def\adr{\right\}}
\def\rar{\rightarrow}
\def\nnb{\nonumber}
\def\la{\langle}
\def\ra{\rangle}
\def\lla{\left<}
\def\rra{\right>}
\def\ba{\begin{array}}
\def\ea{\end{array}}
\def\tep{$B \rar K \ell^+ \ell^-$}
\def\tepm{$B \rar K \mu^+ \mu^-$}
\def\tept{$B \rar K \tau^+ \tau^-$}
\def\ds{\displaystyle}

% ...........................................................

%       the stuff below frames the table

\def\bos{\lower 0.5cm\hbox{{\vrule width 0pt height 1.2cm}}}
\def\boss{\lower 0.35cm\hbox{{\vrule width 0pt height 1.cm}}}
\def\aaa{\lower 0.cm\hbox{{\vrule width 0pt height .7cm}}}
\def\dol{\lower 0.4cm\hbox{{\vrule width 0pt height .5cm}}}

% ...........................................................

\title{ {\Large {\bf The radiative $\Delta \rar N \gamma$ decay
in light cone QCD } } }

\author{\vspace{1cm}\\
{\small T. M. Aliev \thanks
{e-mail: taliev@metu.edu.tr}\,\,,
M. Savc{\i} \thanks
{e-mail: savci@metu.edu.tr}} \\
{\small Physics Department, Middle East Technical University} \\
{\small 06531 Ankara, Turkey} }
\date{}

\begin{titlepage}
\maketitle
\thispagestyle{empty}

\begin{abstract}
\baselineskip  0.7cm
The $g_{\Delta N\gamma}$ coupling for the $\Delta \rar N \gamma$ decay is 
calculated in framework of the 
light cone QCD sum rules and is found to be 
$g_{\Delta N \gamma} = (1.6 \pm 0.2 ) ~GeV^{-1}$. Using this value of 
$g_{\Delta N \gamma}$ we estimate the branching ratio of the
$\Delta^+ \rar N \gamma$ decay, which is in a very good agreement with the
experimental result.   
\end{abstract}

\vspace{1cm}
%PACS numbers: 13.20.He, 11.55.Hx, 12.38.Cy
\end{titlepage}

\section{Introduction}
The extraction of the fundamental parameters of hadrons from experimental
results requires some information about the physics at large distance. 
Unfortunately such an information can not be achieved from the first
principles of a fundamental theory. Indeed QCD, which is believed to be 
the candidate of such an underlying theory of the strong interactions, has
very complicated infrared behavior which makes it impossible to calculate
the properties of of the hadrons starting from a fundamental QCD Lagrangian.    
Therefore for determination of the parameters of hadrons a reliable 
nonperturbative approach is needed. Among all nonperturbative approaches, 
QCD sum rules method which was originally proposed by Shifman, Vainshtein 
and Zakharov \cite{R1} and adopted or extended in many works 
\cite{R2}--\cite{R6}, is particularly a powerful one in studying the 
properties of the low--lying hadrons. In traditional QCD sum rules method 
\cite{R1} the nonperturbative approach is taken into account
through various condensates in the nontrivial QCD vacuum.    

In this work we employ an alternative approach to the traditional sum rules,
namely light cone QCD sum rules method, to study $\Delta \rar N \gamma$
decay coupling constant. 
Light cone sum rules is based on the operator product expansion on
the light cone, which is an expansion over the twists of the operators rather
than dimensions in the traditional sum rules. The main contribution comes
from the lowest twist operator. The matrix elements of the nonlocal
operators sandwiched between a hadronic state and the vacuum defines the
hadronic wane function (more about application of light cone QCD sum rules
can be found in \cite{R7}--\cite{R16} and references therein). 

In general $\Delta \rar N \gamma$ decay is described
by the electric quadrapole E2 and magnetic dipole M1 transition amplitudes. 
However, it is 
well known that the electric quadrapole amplitude is very small compared to 
that of the magnetic dipole amplitude (see \cite{R17} and references therein).
Therefore in this work we consider only the magnetic dipole contribution.  

The coupling constant $g_{\Delta N \gamma}$ is involved in phenomenological
models in investigation of the many reactions of the strong and
electromagnetic interactions and it is expected to be measured more
precisely in the pion photo production experiments at TJNAL (former CEBAF).

The paper is organized as follows: In section 2 the light cone QCD sum rules
for the radiative $\Delta \rar N \gamma$ is presented. Section 3 is devoted
to the numerical analysis and discussion of the results.

\section{The light cone QCD sum rules for $\Delta \rar N \gamma$ decay
constant}

In studying the $\Delta \rar N \gamma$ decay constant we first introduce the
interpolating currents for the $\Delta$ and $N$ baryons \cite{R2}
\bea
\eta_{\Delta^+}^\mu &=& \frac{1}{\sqrt{3}}\, \epsilon^{abc} \left[
\ga u_a^T {\cal C}  \gamma^\mu u_b \dr d_c +
2 \ga u_a^T {\cal C} \gamma^\mu d_b \dr u_c \right]~, \nnb \\
\eta_N &=& \epsilon^{abc} \ga u_a^T {\cal C}  \gamma_\mu u_b \dr
\gamma_5 \gamma^\mu d_c~,
\eea
where $u,~d$ are up and down quark fields, respectively, and
${\cal C}$ is the charge conjugation operator, $a,~b$ and $c$ are the color
indices. Note that this choice of nucleon interpolating currents is not
unique and other choices can be used (see for example \cite{R3}). 

The overlap amplitudes  of the 
interpolating currents with the baryons are defined as
\bea
\lla 0 \vel \eta_N \ver N \rra &=& \lambda_N u_N~, \nnb \\ 
\lla 0 \vel \eta_\Delta^\mu \ver \Delta \rra &=& 
\frac{1}{\sqrt{3}} \lambda_\Delta u_\Delta^\mu ~,
\eea
where $u_\Delta^\mu$ is the Rarita--Schwinger spinor. 

The coupling constant $g_{\Delta N\gamma}$
for the $\Delta \rar N \gamma$ decay is defined as follows
\bea
\lla N \gamma \vel \right.\Delta\rra  = i e g_{\Delta N \gamma} 
\epsilon_{\mu\nu\alpha\beta} \bar u_N 
\gamma^\nu q^\alpha \varepsilon^\beta u_\Delta^\mu~,
\eea
where $\varepsilon_\mu$ and $q_\mu$ are the polarization vector and momentum
of the photon,
respectively, $e$ is the electric charge. 

According to the QCD sum rules ideology, the quantitative estimates of the 
$g_{\Delta N\gamma}$ coupling constant can be obtained by equating the two
different representations of a suitable correlator, written in terms of
hadrons and quark--gluon language. For this purpose we start our analysis by
considering the following correlator
\bea
\Pi(p,q) = \int d^4x \, e^{ipx} 
\lla \gamma(q) \vel \mbox{\rm T}\left\{ \eta_\Delta (0)
\bar \eta_N (x)\right\} \ver 0 \rra~.
\eea
Saturating (4) by $\Delta$ and nucleons and using Eqs. (2) and (3), we get  
for the phenomenological part of the correlator
\bea
i e g_{\Delta N \gamma} \frac{\lambda_N \lambda_\Delta}
{\sqrt{3}(p^2-m_N^2) \left[(p+q)^2-m_\Delta^2\right]}
\Big(\not\!p \left[ \varepsilon_\mu (pq) - (\varepsilon p) q_\mu
\right]\Big) + \mbox{\rm other structures}.
\eea
The main theoretical problem being the calculation of Eq. (4) in QCD. This
problem can be solved in the deep Euclidean region where $p^2$ and $(p+q)^2$
are negative and large. After lengthy calculations, 
at quark level we have obtained the following expression for the correlator
function
\bea
\lefteqn{\Pi(p,q) = 
- \frac{1}{4\sqrt{3}} \int d^4x\,e^{ipx}} \nnb \\
&&\times\Bigg\{ - 4 \lla \gamma(q) \vel \bar d \gamma_5 \gamma_\varphi d
\Big[ \gamma_5 \gamma_\varphi \gamma_\rho \gamma_5 \mbox{\rm Tr}
\ga {\cal S} \gamma_\rho {\cal S}^\prime \gamma_\mu \dr +
2 {\cal S} \gamma_\rho {\cal S}^\prime \gamma_\mu \gamma_5 \gamma_\varphi
\gamma_\rho \gamma_5 \Big] \ver 0 \rra  \\
&&+ 2 \lla \gamma(q) \vel \bar d \sigma_{\alpha\beta}
\Big[ \sigma_{\alpha\beta} \gamma_\rho \gamma_5 \mbox{\rm Tr}
\ga {\cal S} \gamma_\rho {\cal S}^\prime \gamma_\mu \dr +
2 {\cal S} \gamma_\rho {\cal S}^\prime \gamma_\mu \sigma_{\alpha\beta}
\gamma_\rho \gamma_5 \Big]\ver 0 \rra \nnb \\
&&- 4 \lla \gamma(q) \vel \bar u \gamma_5 \gamma_\varphi u 
\Big[ 2 {\cal S} \gamma_\rho \gamma_5 \mbox{\rm Tr}
\ga \gamma_5 \gamma_\varphi \gamma_\rho {\cal S}^\prime \gamma_\mu \dr -
2 {\cal S} \gamma_\rho \gamma_5 \gamma_\varphi \gamma_\mu 
{\cal S} \gamma_\rho \gamma_5 +
2 \gamma_5 \gamma_\varphi \gamma_\rho {\cal S}^\prime \gamma_\mu
{\cal S} \gamma_\rho \gamma_5 \Big]\ver 0 \rra\nnb \\
&&+2 \lla \gamma(q) \vel \bar u \sigma_{\alpha\beta} u
\Big[ 2 {\cal S} \gamma_\rho \gamma_5 \mbox{\rm Tr}
\ga \sigma_{\alpha\beta} \gamma_\rho {\cal S}^\prime \gamma_\mu \dr +
2 {\cal S} \gamma_\rho \sigma_{\alpha\beta} \gamma_\mu
{\cal S} \gamma_\rho \gamma_5 +
2 \sigma_{\alpha\beta} \gamma_\rho {\cal S}^\prime \gamma_\mu
{\cal S} \gamma_\rho \gamma_5 \Big]\ver 0 \rra \Bigg\}~,\nnb
\eea
where ${\cal S}^\prime \equiv {\cal C}{\cal S}{\cal C} =
-{\cal C}{\cal S}{\cal C}^{-1}$ and
$i{\cal S}(x)$ is the full light quark propagator with both
perturbative and nonperturbative contributions
\bea
i{\cal S}(x,0) &=& \lla 0 \vel 
\mbox{\rm T} \left\{ \bar q(x) q(0) \right\} \ver 0 \rra \nnb \\
&=& i \frac{\not\! x}{2 \pi^2 x^4} - \frac{\la \bar q q \ra}{12} -
\frac{x^2}{192} m_0^2 \la \bar q q \ra \nnb \\
&-& i g_s \frac{1}{16 \pi^2} \int_0^1 du \Bigg\{
\frac{\not\! x}{x^2} \sigma_{\alpha\beta}  G^{\alpha\beta}(ux)
- 4 i u \frac{x_\mu}{x^2} G^{\mu\nu}(ux)\gamma_\nu \Bigg\} 
+ \cdots 
\eea
It follows from Eq. (6) that, in order to calculate the correlator function
$\Pi$ in QCD, the matrix elements
$\lla \gamma(q) \vel \bar q \gamma_\alpha \gamma_5 q \ver 0 \rra$ and 
$\lla \gamma(q) \vel \bar q \sigma_{\alpha\beta} q \ver 0 \rra$ are needed.
These matrix elements are defined in terms of the photon wave functions as
follows \cite{R18}--\cite{R20}
\bea
\lla \gamma(q) \vel \bar q \gamma_\alpha \gamma_5 q \ver 0 \rra &=& 
\frac{f}{4} e_q e \epsilon_{\alpha\beta\rho\sigma} \varepsilon^\beta q^\rho x^\sigma
\int_0^1 du \, e^{iuqx} \psi(u) ~, \nnb \\ \nnb \\
\lla \gamma(q) \vel \bar q \sigma_{\alpha\beta} q \ver 0 \rra &=&
i e_q e \la \bar q q \ra \int_0^1 du \, e^{iuqx} \nnb \\
&\times& \Bigg\{
\ga \varepsilon_\alpha q_\beta - \varepsilon_\beta q_\alpha \dr \Big[ \chi \phi (u) +
x^2 \left[ g_1 (u) - g_2 (u) \right] \Big] \nnb \\
&+& \Big[ qx \ga \varepsilon_\alpha x_\beta - \varepsilon_\beta x_\alpha \dr +
\varepsilon x
\ga x_\alpha q_\beta - x_\beta q_\alpha \dr \Big] g_2 (u) \Bigg\}~.
\eea
where the parameter $\chi$ is the magnetic susceptibility of the quark 
condensate and $e_q$ is the quark charge. In further analysis the path 
ordered gauge factor 
\bea
{\cal P} exp \ga \int_0^1 du x^\mu A_\mu (ux) \dr~, \nnb
\eea
is omitted since in the fixed point gauge $x^\mu A_\mu=0$.
The functions $\phi(u)$ and $\psi(u)$ in Eq. (8) are the leading 
twist photon functions, while $g_1(u)$ and $g_2(u)$ are the twist--4   
functions. Using Eqs. (6), (7) and (8), and performing Fourier transform for
the structure $\left[ \varepsilon_\mu (qp) - (\varepsilon p) q_\mu \right]$,
we get the following result
\bea
\Pi &=& -\frac{i}{\pi^2} \frac{\la \bar q q \ra}{4\sqrt{3}} \int_0^1 du \nnb
\\ 
&\times&\Bigg\{\ga e_u - e_d \dr \Bigg[\frac{1}{3} \chi 
\phi(u) \mbox{\rm ln}(-P^2)
- \Big[ 4 g_1(u)+2 g_2(u) \Big] \frac{1}{P^2} +
\frac{2 \pi^2}{3 P^4} f \psi(u) + \frac{\pi^2}{3 P^6} f m_0^2 \psi \Bigg]
\nnb \\
&+& \frac{1}{3 P^6} g_2(u) \la g^2 G^2 \ra \ga - 7 e_u + 3 e_d \dr 
-\frac{1}{12} e_d \la g^2 G^2 \ra \chi \phi(u) \frac{1}{P^4} -
\frac{2}{3} g_1(u) e_d \la g^2 G^2 \ra \frac{1}{P^6} \Bigg\}~,
\eea
where $P = p + qu$.
 
As has been noted already, the QCD sum rule is obtained as usual by equating
the hadronic representation of the correlator (4) with the result of the
QCD calculation. In order to take into account the contributions of the higher 
states we invoke
the quark--hadron duality prescription, i.e., above certain
thresholds in $s_1$ and $s_2$, the double spectral density $\rho(s_1,s_2)$
for the higher states and continuum coincides with the 
spectral density calculated in QCD.

After performing double Borel transformation with respect to
the variables $p^2$ and $(p+q)^2$ in Eqs. (4) and (9) to suppress the higher
states, we finally get the
following sum rules for the $\Delta N\gamma$ coupling constant
\bea
g_{\Delta N\gamma} \lambda_\Delta \lambda_N &=& - \frac{1}{4} e^{m^2/M^2}
\frac{\la \bar q q \ra}{\pi^2} \Bigg\{ \ga e_u - e_d \dr \Bigg[-\frac{1}{3} 
\chi \phi(u_0) M^4 f_1(s_0/M^2) \nnb \\
&+& \Big[4 g_1(u_0) + 2 g_2(u_0) \Big] M^2 f_0(s_0/M^2) +
\frac{2 \pi^2}{3} f \psi(u_0) \ga 1 - \frac{m_0^2}{4 M^2} \dr \Bigg] \nnb \\
&+&\Bigg[\frac{1}{6 M^2}\ga - 7 e_u + 3 e_d \dr g_2(u_0) + 
\frac{1}{3 M^2} g_1(u_0) - \frac{e_d}{12} \chi \phi(u_0) \Bigg] 
\la g^2 G^2 \ra \Bigg\}~,
\eea
where the function 
\bea
f_n(x)=1-e^{-x}\sum_{k=0}^n \frac{(x)^k}{k!}~, \nnb
\eea
is the factor used to
subtract the continuum, $s_0$ is the continuum threshold and
\bea   
u_0 = \frac{M_2^2}{M_1^2 + M_2^2}~,
~~~~M^2 = \frac{M_1^2 M_2^2}{M_1^2 + M_2^2}~, \nnb
\eea
where $M_1^2$ and $M_2^2$ are the Borel parameters. Since masses of the proton,
and $\Delta^+$ are very close to each other, we can choose 
$M_1^2$ and $M_2^2$ to be equal to each other, i.e., $M_1^2 = M_2^2 =2 M^2$,
from which it follows that $u_0=1/2$.  
\section{Numerical results}
It follows from eq. (10) that the main input parameters of the sum rules are
photon wave functions. It was shown in \cite{R7,R8} that the leading photon
wave functions receives only small corrections from the higher conformal
spin, so they do not deviate much from the asymptotic form. Following
\cite{R18}--\cite{R20}, we shall use for the photon wave function
\bea
\phi(u) &=& 6 u \bar u~, \nnb \\
\psi(u) &=& 1 ~,  \nnb \\
g_1(u) &=& - \frac{1}{8} \bar u (3 -u)~,  \nnb \\
g_2(u) &=& - \frac{1}{4} \bar u^2~, \nnb
\eea
where $\bar u = 1-u$. The values of the other input parameters we have used: 
$f = 0.028~GeV^2$ and $\chi = -4.4~GeV^2$ \cite{R20} at the scale
$\mu=1~GeV$, $\la g^2 G^2 \ra = 0.474~GeV^2$ and for the continuum threshold
$s_0$ we have chosen two different values, i.e., $s_0=2.8~GeV^2$ and
$s_0=3~GeV^2$. Having fixed the input parameters, one must find the range of
values of $M^2$ for which the sum rule (10) is reliable. The lowest value
of $M^2$ is usually determined by the condition that the terms proportional
to the highest inverse power of the Borel parameter stay reasonably small.
The upper limit is determined by demanding that the continuum and and higher
state contribution does not get too large, say less than $30\%$ of the
leading twist contributions. Both conditions are satisfied in the interval
$1~GeV^2 \le M^2 \le 1.5~GeV^2$. The dependence of the right side of 
Eq. (10) on $M^2$ is shown in Fig. (1). It follows from this figure in that
the above--mentioned working region of $M^2$ the sum rule is quite stable.
From this figure one can directly predict that 
\bea 
g_{\Delta N\gamma} \lambda_\Delta \lambda_N = (0.0020 \pm 0.0004) ~ GeV^5.
\eea
Dividing this product of couplings by the residues of proton and $\Delta$
currents $\lambda_N$ and $\lambda_\Delta$, that were calculated in the 
analysis of mass sum rules for baryons
\cite{R2} (see also \cite{R3} and \cite{R4})
\bea
g_{\Delta N\gamma}=  \ga 1.6 \pm 0.2 \dr ~GeV^{-1}.
\eea
This prediction of the coupling constant permits us to estimate the width of
the $\Delta \rar N\gamma$ decay. Using the matrix element for the 
$\Delta \rar N\gamma$ transition (see eq. (3)), we get for the decay width
\bea
\Gamma = \frac{\alpha}{4 m_\Delta^3} g_{\Delta N\gamma}^2 
\ga m_\Delta^2-m_N^2 \dr^3 
\left[ 1 + \frac{1}{3m_\Delta^2} \ga m_\Delta-m_N \dr^2 \right]~,
\eea
where $\alpha$ is the fine structure constant, $m_\Delta$ and $m_N$ are the
masses of $\Delta$ and nucleons, respectively. Using the predicted value of
$g_{\Delta N\gamma}$ in Eq. (12), the result we get for the decay width is
\bea
\Gamma \simeq 0.65 \ga 1 \pm 0.25 \dr ~MeV~, \nnb
\eea
and for the branching ratio of this channel we have (for the total decay
width we have used $\Gamma_{tot} = 113~MeV$ \cite{R22})
\bea
{\cal B}(\Delta \rar N \gamma) = \frac{\Gamma}{\Gamma_{tot}} =
0.0058 ~. \nnb
\eea
This prediction is in a very good agreement with the experimental results,
i.e., $0.52\% \le {\cal B} \le 0.60 \%$ \cite{R22}.

In summary, we have calculated $\Delta N\gamma$ coupling using the light
cone QCD sum rules. Our prediction on the branching ratio is in a good
agreement with the experimental results.
\section*{Acknowledgments}
We are grateful to M. P. Rekalo for useful discussions.
\newpage
\begin{figure}   
\vskip 1.5cm
    \includegraphics{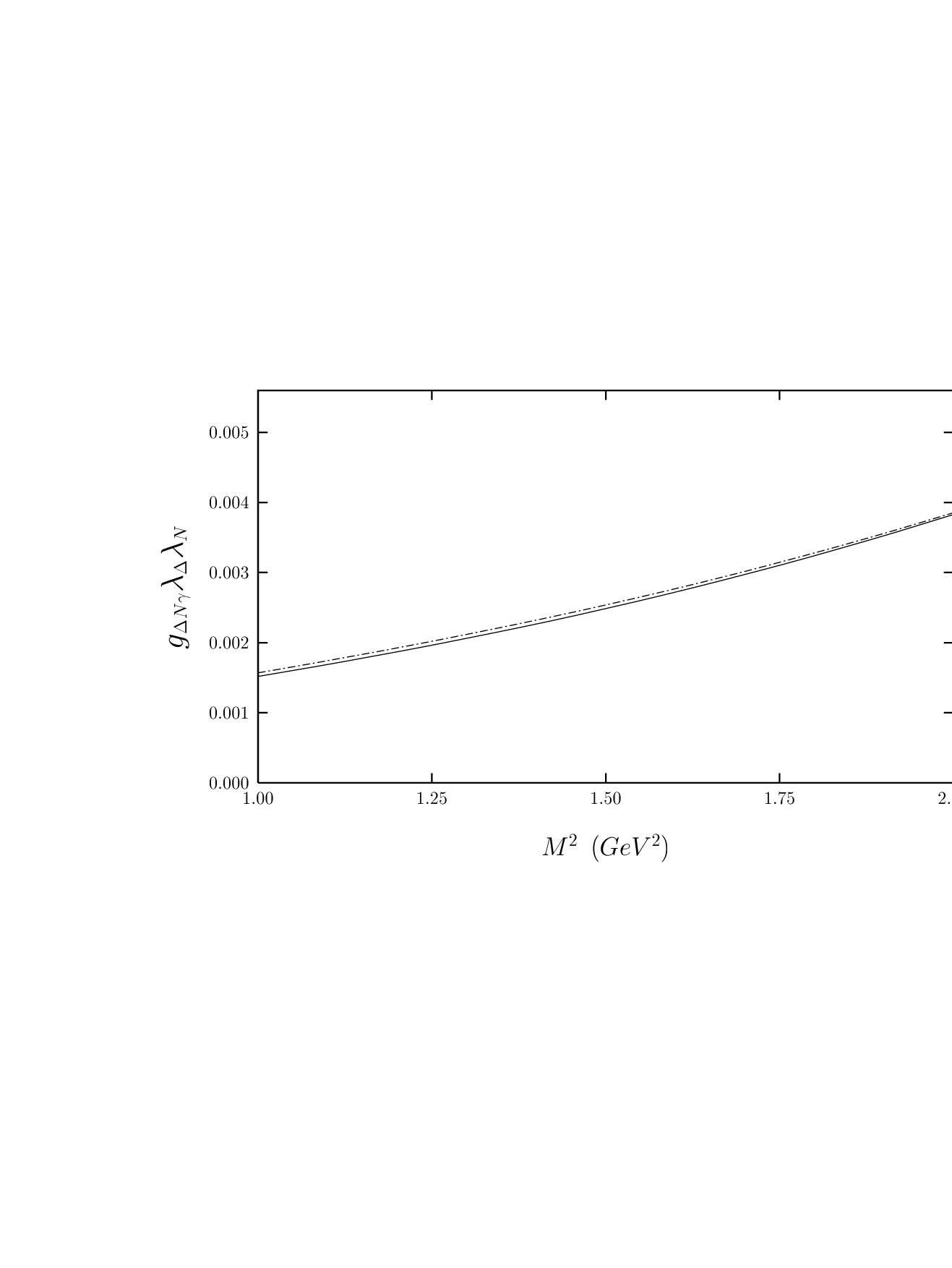}
\vskip 6.5cm  
\caption{}
\vskip 10cm
\end{figure}  
\newpage
\section*{Figure captions}   
{\bf Fig. 1} The dependence of $g_{\Delta N\gamma} \lambda_\Delta \lambda_N$
on the Borel parameter $M^2$. In this figure the solid and dash--dotted 
lines correspond to the threshold values $s_0=2.8~GeV^2$ and $s_0=3.0~GeV^2$, 
respectively.


\newpage
\begin{thebibliography}{99}

\bibitem{R1} M. A. Shifman, A. I. Vainshtein, and V. I. Zakharov,
{\it Nucl. Phys.} {\bf B147} (1979) 385; {\bf B147} (1979) 448;
{\bf B147} (1979) 519.

\bibitem{R2} B. L. Ioffe,
{\it Nucl. Phys.} {\bf B188} (1981) 317;
Errata, {\it Nucl. Phys.} {\bf B191} (1981) 591.

\bibitem{R3} V. Chung, H. G. Dosch, M. Kremer and D. Schall,
{\it Phys. Lett.} {\bf B102} (1981) 175;\\
{\it Nucl. Phys.} {\bf B197} (1982) 55.

\bibitem{R4} L. I. Reinders, H. R. Rubinstein and S. Yazaki,
{\it Phys. Rep.} {\bf 127C} (1985) 1.

\bibitem{R5} V. M. Belyaev and B. L. Ioffe,
{\it Sov. J. JETP} {\bf 56} (1982) 493;
B. L. Ioffe and A. V. Smilga,
{\it Phys. Lett.} {\bf B114} (1982) 353;
{\it Nucl. Phys.} {\bf B232} (1984) 109.

\bibitem{R6} I. I. Balitsky, A. V. Yung,
{\it Phys. Lett.} {\bf B129} (1983) 328. 

\bibitem{R7} I. I. Balitsky, V. M. Braun, A. V. Kolesnichenko,
{\it Nucl. Phys.} {\bf B312} (1989) 509.   

\bibitem{R8} V. M. Braun, I. E. Filyanov,
{\it Z. Phys.} {\bf C48} (1990) 239; {\it Z. Phys.} {\bf C44} (1989) 157.

\bibitem{R9} V. L. Chernyak and A. R. Zhitnitsky,
{\it JETP Lett.} {\bf 25} (1977) 510;\\
A. V. Efremov and a. V. Radyushkin,
{\it Phys. Lett.} {\bf 94B} (1980) 245;\\
G. P. Lepage and S. J. Brodsky,
{\it Phys. Lett.} {\bf 87B} (1979) 359.

\bibitem{R10} V. L. Chernyak and A. R. Zhitnitsky,
{\it Phys. Rep.} {\bf 112} (1984) 173.

\bibitem{R11} V. M. Braun,
In proc. "Rostock 1997, Progress in heavy quark physics" 105-118, 1997,
prep. hep-ph/9801222.

\bibitem{R12} V. L. Chernyak and A. R. Zhitnitsky,
{\it Nucl. Phys.} {\bf B345} (1990) 137.

\bibitem{R13} V. M. Belyaev, V. M. Braun, A. Khodjamirian and R. R\"{u}ckl,\\
{\it Phys. Rev.} {\bf D51} (1995) 6177.

\bibitem{R14} P. Ball, V. M. Braun,
{\it  Phys. Rev} {\bf D55} (1997) 5561.

\bibitem{R15} P. Ball, V. M. Braun,
{\it  Phys. Rev} {\bf D58} (1998) 094016.

\bibitem{R16} T. M. Aliev, M. Savc{\i}, A. {\"O}zpineci,
{\it Phys. Rev.} {\bf D56} (1997) 4260.

\bibitem{R17} M. N. Butler, M. Savage, R. Springer,
{\it Phys. Lett.} {\bf B304} (1993) 353.

\bibitem{R18} G. Eilam, I. Halperin and R. R. Mendel,
{\it Phys. Lett.} {\bf B361} (1995) 137.

\bibitem{R19} A. Ali, V. M. Braun,
{\it Phys. Lett.} {\bf B359} (1995) 223.

\bibitem{R20} A. Khodjamirian, G. Stoll and D. Wyler,
{\it Phys. Lett.} {\bf B358} (1995) 129.

\bibitem{R21} V. M. Belyaev, Y. I. Kogan,
{\it Yad. Phys.} {\bf 40} (1984) 1035;\\
I. I. Balitskii, A. V. Kolesnichenko,
{\it Yad. Phys.} {\bf 41} (1985) 282.

\bibitem{R22} Particle Data Group, C. Caso {\it et al.},
{\it Eur. Phys. J.} {\bf C3} (1998) 1.

\end{thebibliography}
\end{document}